\documentstyle[tighten,aps,psfig,epsf,axodraw,floats]{revtex}

\newcommand{\be}{\begin{equation}}
\newcommand{\ee}{\end{equation}}
\newcommand{\ba}{\begin{eqnarray}}
\newcommand{\ea}{\end{eqnarray}}

\newcommand{\ep}{\epsilon}

\begin{document}

\twocolumn

\draft

\title{
\[ \vspace{-2cm} \]
\noindent\hfill\hbox{\rm  SLAC-PUB-8844} \vskip 1pt
\noindent\hfill\hbox{\rm hep-ph/0105267} \vskip 10pt
On the theoretical uncertainties 
in the muon anomalous magnetic moment
}

\author{Kirill Melnikov\thanks{
e-mail:  melnikov@slac.stanford.edu}}
\address{Stanford Linear Accelerator Center\\
Stanford University, Stanford, CA 94309}

\maketitle

\begin{abstract}
I present a fairly detailed discussion of  various contributions 
to the anomalous magnetic moment of the muon $a_\mu$. 
I try to give an unbiased evaluation  of the validity of 
the SM prediction for this quantity and to point out 
some delicate issues involved in its calculation. 
I conclude that the theory uncertainties in the 
SM prediction for the muon anomalous magnetic moment are underestimated and 
a great deal of work will be required to reduce these 
uncertainties to the level required by experiment.
\end{abstract}

\section{Introduction}

Recently, the E821 experiment at Brookhaven National Laboratory 
reported a new value 
of the anomalous magnetic moment of the muon, based on the 
$\mu^+$ data collected through 1999. Their result. 
averaged with previous measurements, leads to 
a new world average \cite{g-2recent}, 
\be
a_\mu^{\rm exp} = 116~592~020(150) \times 10^{-11},
\ee
that is claimed to be $2.6\sigma$ away from the Standard Model prediction:
\be
a_\mu^{\rm th} = 116~591~597(67) \times 10^{-11}.
\label{thp}
\ee
There are numerous theoretical papers  devoted to the 
interpretation of  this apparent discrepancy as the direct signal of  
physics beyond the Standard Model (SM)  (see  \cite{home} for examples).

Because of the potential importance of this result and because 
of the subtlety of certain of the SM contributions, it is important 
to carefully review the corresponding calculations.
Such a  review has been presented 
recently by A.~Czarnecki and W.~Marciano \cite{CM}; however, 
in my opinion, more attention should be paid to certain 
aspects of the problem.

This article is organized in  the following way.  First, I briefly 
describe all the contributions to the SM value  for $a_\mu$ to remind 
the reader   what went into the  theory result quoted by the g-2 
collaboration.  I then concentrate on the hadronic contribution to 
photon vacuum polarization and  discuss its evaluation based on 
both $\tau$ and $e^+e^-$ data.  After that I  describe the 
hadronic light-by-light scattering contribution.

There are three major questions I would like to address in this article:
1) Is there a g-2 crisis? 2) What should 
be done to make a solid  case for the crisis? 3) What are the odds that 
there {\it will be} a crisis after E821 reaches its projected accuracy 
of $40 \times 10^{-11}$. My analysis indicates that the theory uncertainties 
in $a_\mu$ are larger than indicated in Eq.(\ref{thp}) and that a great 
deal of sophisticated work will be required to reduce them.

I hope the information presented here will be useful both, for 
a person who is about to post a New Physics paper on the 
muon anomalous magnetic moment at the LANL archive, and for a person who wants 
to understand the down-to-earth physics issues involved in 
the calculation of $a_\mu$.  

Finally, I should apologize to the many experts on the anomalous magnetic 
moment of the muon who have done the hard  work to achieve the 
accuracy of the SM prediction that we now have. 
Although I have not done 
any of the original theoretical work, 
a fresh look at these important calculations might be useful. I also 
hope that there are some points  in my discussion 
that might be of interest even for the  experts working in the field. 

\section{What theoretical input is in the $2.6~\sigma$ discrepancy?}

Let us first make clear to ourselves what went into the theoretical 
number that is $2.6~\sigma$ away from the result of the 
g-2 Collaboration. 
It is a common practice to write the muon anomalous magnetic moment 
as the sum of the QED, weak and hadronic contributions:
\be
a_{\mu}^{\rm th} = a_\mu^{\rm QED} + a_\mu^{\rm weak} + a_\mu^{\rm hadr},
\ee
for which the following values have been used by the g-2 Collaboration:
\ba
&& a_\mu^{\rm QED} = 116~584~705.7(2.9) \times 10^{-11},
\label{amuqed} \\
&& a_\mu^{\rm weak} = 152(4) \times 10^{-11},
 \\
&&a_\mu^{\rm had} = 6739(67) \times 10^{-11}.
\ea
The first thing to notice here is the one per cent accuracy of the 
hadronic contribution to $a_\mu$ and I will discuss how trustworthy 
it is  in detail in the rest of the paper. Before that, however, 
I would like to comment on the QED and weak contributions.

The pure QED contribution (which only includes muon, electron
and $\tau$) is known to five loops\footnote{More precisely, 
four loops are computed, the ${\cal O}(\alpha^5)$ contribution is 
estimated.}. Note that  the value of the fine structure constant 
obtained from the electron g-2 has been employed here but 
the result reported in Eq.(\ref{amuqed}) does not change
significantly if more conservative values 
for  the fine structure constant (for example, as obtained from 
the quantum Hall effect) are used to evaluate it.

At the current level of precision
the five loop QED contribution is not yet relevant; roughly, it gives 
$5 \times 10^{-11}$. The four loop contribution, on the other 
hand is quite relevant; its contribution  to Eq.(\ref{amuqed})
is $\sim 400 \times 10^{-11}$.  I think it is appropriate  
to ask how well the four loop contribution is known.
The answer to this question is not quite simple. The complete 
four loop calculation has been performed by a single group
and has never been checked by an independent calculation. 
To see that there is a potential 
problem here, one can look at how the central value of the four loop 
contribution   evolved in time . In \cite{kin1} it was reported as 
$140(6)(\alpha/\pi)^4$ and  was later revised \cite{kin2} to 
$126.92(41)(\alpha/\pi)^4$. The shift, 
$\sim 40 \times 10^{-11}$,  is a non-negligible change on the level 
of precision $100 \times 10^{-11}$. 
On the other hand, about $90$ per cent of the full ${\cal O}(\alpha^4)$
result comes from a simple class of diagrams where 
the electron vacuum polarization is inserted  into 
one of three photons in ${\cal O}(\alpha^3)$ electron light-by-light
scattering diagrams. This particular contribution has been evaluated 
several times  with the result $116(\alpha/\pi)^4$ (see e.g. \cite{samuel}). 
Unfortunately, this fact alone does not tell us much. 
It is clear 
that 1) mistakes, in general, can happen and 2) we are talking here 
about one of the most heroic and sophisticated calculations {\it ever} 
performed in perturbative 
quantum field theory.
For the rest of this article  we will  assume that  
$a_\mu^{\rm QED}$, as given in 
Eq.(\ref{amuqed}), is correct but it is  important to keep in mind that 
even in the QED part of $a_\mu$ there is, potentially, some work to be done 
if we want to gain complete confidence in the SM prediction for this quantity.

The  weak contribution to $a_\mu$ is currently known up to two loops
\cite{oneloop,twoloop}. The result is:
\be
a_\mu^{\rm weak} = \left ( 195 - 43(4)  \right ) \times 10^{-11} = 
152(4) \times 10^{-11},
\ee
where the one-loop and the two-loop contributions are displayed 
separately. The two-loop correction seems to be too large for a ``normal''
electroweak correction. 
The  reason for such a big second order effect is that 
large logarithms $\log (m_W/m_f)$ where $m_f$ is  the mass 
of a light fermion (muon, electron or any of the quarks)  
appear in the two loop diagrams for the first time \cite{kuraev1}. 
These logarithms make up the bulk
of the second order correction and they can be summed up \cite{peres}
using renormalization group techniques. This has been done \cite{peres} 
and it did not change the value  quoted above significantly. 
So, the weak corrections seem to be well established.

\begin{figure}
\begin{center}
\hfill
\begin{picture}(120,40)(0,0)
 \SetScale{.8}
 \SetWidth{1.5}
 \Line(20,20)(110,20)
 \SetWidth{0.8}
  \ArrowLine(25,20)(42,20)
 \ArrowLine(42,20)(62,20)
 \ArrowLine(67,20)(87,20)
 \ArrowLine(80,20)(96,20)
 \ArrowLine(100,20)(110,20)
 \SetWidth{2}
  \Oval(70,40)(7,11)(0)
 \SetWidth{0.5}
  \Line(65,45)(70,32)
  \Line(70,45)(75,34)
  \Line(75,45)(80,36)
 \PhotonArc(70,14)(27,13,70){1.6}{6}
 \PhotonArc(70,14)(27,110,167){1.6}{6}
  \Photon(70,20)(70,0){1.8}{4}
 \Text(20,-2)[t]{(a)}
\end{picture}
\hfill
\begin{picture}(120,40)(0,0)
 \SetScale{.8}
 \SetWidth{1.5}
 \Line(20,20)(110,20)
 \SetWidth{0.8}
  \ArrowLine(25,20)(42,20)
 \ArrowLine(42,20)(62,20)
 \ArrowLine(67,20)(87,20)
 \ArrowLine(80,20)(96,20)
 \ArrowLine(100,20)(110,20)
 \SetWidth{2}
  \Oval(70,40)(7,11)(0)
 \SetWidth{0.5}
  \Line(65,45)(70,32)
  \Line(70,45)(75,34)
  \Line(75,45)(80,36)
 \PhotonArc(45,24)(14,197,343){1.6}{7.5}
 \PhotonArc(70,14)(27,13,70){1.6}{6}
 \PhotonArc(70,14)(27,110,167){1.6}{6}
  \Photon(70,20)(70,0){1.8}{4}
 \Text(20,-2)[t]{(b)}
\end{picture}
\hfill\null\\
\hfill
\begin{picture}(120,60)(0,0)
 \SetScale{.8}
 \SetWidth{1.5}
 \Line(20,20)(110,20)
 \SetWidth{0.8}
  \ArrowLine(25,20)(42,20)
 \ArrowLine(42,20)(62,20)
 \ArrowLine(67,20)(87,20)
 \ArrowLine(80,20)(96,20)
 \ArrowLine(100,20)(110,20)
 \SetWidth{2}
  \Oval(70,40)(7,11)(0)
 \SetWidth{0.5}
  \Line(65,45)(70,32)
  \Line(70,45)(75,34)
  \Line(75,45)(80,36)

 \PhotonArc(70,44)(10,11,167){1.6}{6}
 \PhotonArc(70,14)(27,13,70){1.6}{6}
 \PhotonArc(70,14)(27,110,167){1.6}{6}
  \Photon(70,20)(70,0){1.8}{4}
 \Text(20,-2)[t]{(c)}
\end{picture}
\hfill
\begin{picture}(120,60)(0,0)
    \SetScale{.8}
 \SetWidth{1.5}
 \Line(5,30)(130,30)
  \SetWidth{1.3}
  \Oval(70,9)(7,11)(0)
 \SetWidth{0.8}
 \ArrowLine(15,30)(28,30)
 \ArrowLine(42,30)(62,30)
 \ArrowLine(83,30)(99,30)
 \ArrowLine(115,30)(120,30)

  \ArrowLine(61.5,14)(66.5,16)
  \ArrowLine(74,15.5)(80,13)
   \ArrowLine(79,5.0)(73,2)
   \ArrowLine(65,2.7)(60,5.5)

 \SetWidth{0.5}
  \Line(60,15)(66,3)
  \Line(65,15)(72,4)
  \Line(70,15)(77,5)

  \Photon(70,30)(70,16){1.8}{3}
 \PhotonArc(59,36)(25,195,270){1.6}{7.5}
 \PhotonArc(81,36)(25,270,345){1.6}{7.5}
  \Photon(70,2)(70,-10){1.8}{2.5}
 \Text(20,-2)[t]{(d)}
\SetScale{1}
\end{picture}
\hfill\null\\
\vglue 18pt
\end{center}
\caption{Examples of hadronic contributions to g-2. a) Leading order 
hadronic vacuum polarization diagram; b) example of the next-to-leading 
order hadronic vacuum polarization diagram;
c) the diagram that is {\it not} 
considered as part of the next-to-leading order hadronic vacuum polarization 
diagrams in the usual nomenclature; d) hadronic light-by-light.}
\label{fig:examplediagrams}
\end{figure}
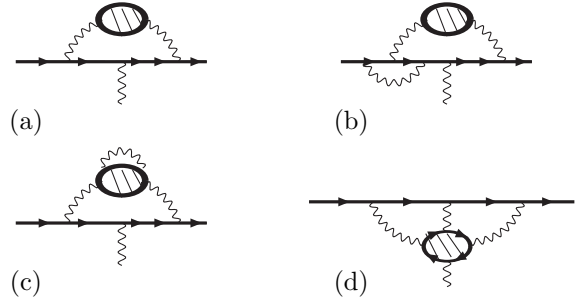

Finally,  hadronic contributions to the muon anomalous magnetic 
moment are usually separated into hadronic vacuum polarization 
(which is further separated into the leading and the next-to-leading 
order pieces) and hadronic  light-by-light scattering 
contributions (see Fig.1). In Ref. \cite{g-2recent}, the following results 
have been used for these contributions:
\ba
&& a_\mu^{\rm hadr}({\rm lo~v.p.}) = 6924(62) \times 10^{-11}
~~~~~\cite{davier},\label{vplo} \\
&& a_\mu^{\rm hadr}({\rm nlo~v.p.}) = -100(6) \times 10^{-11}
~~~~\cite{krause}, \label{vpnlo} \\
&& a_\mu^{\rm hadr}({\rm lbl}) = -85(25) \times 10^{-11}
~~~~~~~~~~\cite{klbl,bijnes}. \label{hlbl}
\ea

The theory prediction, Eq.(\ref{thp}), is obtained by 
taking the sum of the QED, weak and hadronic contributions and 
{\it adding their errors in  quadratures}:
\be
a_\mu^{\rm th} = 116~591~597(67) \times 10^{-11},
\ee
which leads to 
$$
\left [ a_\mu^{\rm exp} - a_\mu^{\rm th} \right ]\times 10^{11} 
= 426 \pm 150 |_{\rm exp} \pm 67|_{\rm th} \equiv 426 \pm 165,
$$
where at the last step the theoretical and experimental errors 
are combined in the quadratures. This difference is interpreted 
as the $2.6~\sigma$ deviation  in \cite{g-2recent}.

Some comments about this result and its interpretation can be 
made immediately. 
First of all,  the dominant source of the theory error is  
the hadronic contribution, in particular the vacuum polarization 
(later I will discuss the hadronic light-by-light scattering result in detail).
A glance at the numbers in Eq.(\ref{vplo}) shows clearly that one needs 
to control the calculation of hadronic vacuum polarization 
at the {\it one per cent level}. 
Hadronic vacuum polarization is  derived by using the dispersion 
representation for the photon 
propagator which relates the hadronic vacuum polarization correction 
to $a_\mu$ and the annihilation cross section $e^+e^- \to {\rm hadrons}$:
\be
a_\mu^{\rm hadr}({\rm lo~v.p.}) = 
\frac {1}{4\pi^3} \int \limits_{4 m_\pi^2}^{\infty} 
{\rm d}s~K(s) \sigma^{0}(s)_{e^+e^- \to {\rm had}}.
\label{eq7}
\ee
For large $s$ the function $K(s)$ behaves, to a good approximation, 
as $m_\mu^2/s$ and  for this reason the contribution of the low $s$ region 
is dominant and any ``first principles'' calculation becomes impossible. 
This forces one to rely on experimental data  
to evaluate $a_\mu^{\rm hadr}({\rm lo~v.p.})$ and for this reason
it is quite  important to know  exactly where the number  
in Eq.(\ref{vplo}) comes from. 
In fact this number is taken from the calculation \cite{davier} 
based, in addition 
to $e^+e^- \to {\rm hadrons}$, on 1) using the data on $\tau$ decays 
supplemented by the conserved vector current (CVC) hypothesis and 
isospin symmetry; 
2) sophisticated machinery of the finite energy QCD sum rules
designed specifically to minimize the errors; 3) application 
of perturbative QCD down to energy scales of about $1.8~{\rm GeV}$ 
($J/\psi$ and $\Upsilon$ families are treated separately, using experimental
input). It is important to notice that   the natural scale 
for both CVC and isospin symmetry violations is the one per cent; 
the smallness of the  error of the result in Eq.(\ref{vplo}) 
is the consequence of  both, the quality of the $\tau$ data {\it and} 
the use of pQCD down to rather low energy energies.

My second comment is that the $2.6~\sigma$ deviation 
only appears if all the errors are combined in quadratures. 
If one  combines all the theory errors linearly, one 
ends up in a somewhat different situation:
\be
\left [ a_\mu^{\rm exp} - a_\mu^{\rm th} \right ]\times 10^{11} 
= 426 \pm 150 |_{\rm exp} \pm 100|_{\rm th}.
\label{lin}
\ee
I would like to stress that it is not at all clear how these numbers 
should be combined  to get the final error since it is a bit too 
strong an assumption to assign a Gaussian distribution to the theory 
error. It is true that it is unclear how to interpret 
Eq.(\ref{lin}) in terms of standard deviations; on the other hand 
it is equally 
unclear that this should 
be done since a glance at the content of the hep-ph archive 
through recent months is probably a good illustration of the fact that 
too many people take the word ``standard deviation'' too literally.

\section{Hadronic vacuum polarization and  the $\tau$ data}

The use of the $\tau$ data for the evaluation of $a_\mu({\rm lo~v.p})$ 
is related 
to the fact that the integration over $s$ in Eq.(\ref{eq7})  
saturates at $\sqrt{s} < 2~{\rm GeV}$:
about $70$ per cent of the total hadronic contribution 
comes from  the two pion final state at energies as low as 
$\sqrt{s} < 1~{\rm GeV}$, and about $90$ per cent from 
the energy region $\sqrt{s} \le 2~{\rm GeV}$. 
The accuracy of the data available until rather 
recently from $e^+e^-$ machines was not quite adequate.  
It then looked natural to combine them with the  
data on $\tau$ decays obtained by the ALEPH
collaboration
to improve the calculation of $a^{\rm hard}_\mu$.

As I have already mentioned,  the essential {\it theoretical} input 
one brings in with the $\tau$ data is 
the CVC hypothesis and the isospin symmetry. 
We know that, generically, these are rather good symmetries. 
We also know that these symmetries should be violated at the 
one per cent level (by e.g. electromagnetism or $(m_\pi/m_\rho)^2$) 
and therefore  the real question here is in how well these violations can be 
controlled.

Both isospin and CVC are violated by the mass difference of the 
up and down quarks and also by the QED corrections. Let us 
discuss the transition from the $\tau$ data 
to $e^+e^- \to {\rm hadrons}$ in some detail to expose potential 
problems here. 

Imagine that we have a perfect measurement of the 
$\tau \to \nu_\tau \pi^0 \pi^-$
branching ratio. Starting from there, one 
identifies several effects that  might affect the transition 
to $e^+e^- \to hadrons$. They are:
1) the short distance QED corrections in $\tau$ decays; 
2) the difference in the masses of charged and neutral pions 
and $m_u-m_d \ne 0$ effects; 3) 
the difference in the decay widths and the masses 
of neutral and charged $\rho$ mesons; 
4) the long-distance QED radiative corrections in $\tau$ decays.

Let us discuss these effects step by step. The short-distance QED corrections 
are Wilson coefficients of the four-fermion operators that describe $\tau$ 
decays; they are generated by  exchanges of photons with the virtualities 
$m_\tau^2 \ll k^2 \ll m_W^2$. Because of this, the short-distance 
corrections are universal in a sense that, for  
a given four-fermion operator, they do not depend on subtle 
details of hadronic final state. Note, however, that due to the difference 
in  relevant four-fermion operators for leptonic and hadronic $\tau$ decays, 
these short distance corrections are absent in 
$\tau \to \nu_\tau + {\rm leptons}$. The short-distance 
QED Wilson coefficient  is \cite{ms}:
\be
S_{\rm ew} = \left ( 1 + \frac{\alpha}{\pi} \log \frac{m_W}{\mu} \right )
\approx 1.009,
\label{wils}
\ee
where $\mu$ is an arbitrary parameter; the numerical value corresponds 
to $\mu = m_\tau$. Since there is no similar renormalization factor 
in $e^+e^-$ annihilation, the relation between $\tau$ decay width 
and the $e^+e^-$ annihilation cross section reads, schematically:
\be
\Gamma(\tau^- \to \nu_\tau \pi^0 \pi^-) \approx S_{\rm ew}^2 
\sigma(e^+e^- \to {\rm hadrons}).
\ee
The renormalization factor $S_{\rm ew}$ is taken into 
account when the $\tau$ data is used to predict $e^+e^- \to {\rm hadrons}$
\cite{davier}; numerically, as can be seen from the numbers above, 
it amounts to renormalizing the $\tau$ data by about 
$-2$ per cent.

The second effect is the difference in the masses of charged and neutral 
pions. Since the pions are produced in the ${\rm P}$-wave, both in 
$\tau^- \to \nu_\tau \pi^0 \pi^-$ and in $e^+e^- \to \pi^+ \pi^-$, 
the rates are proportional to the third power of the velocity of the pions.
In this case, the relation between the $\tau$ decay rate and the 
$e^+e^-$ cross section is:
$$
\Gamma(\tau^- \to \nu_\tau \pi^0 \pi^-) \approx 
\frac{\beta_{\pi_0 \pi^-}^3}{\beta_{\pi^+ \pi^-}^3}
\sigma(e^+e^- \to {\rm hadrons}).
$$
Since the mass difference of charged and neutral pions is relatively large, 
this correction turns out to be in a few per cent range in the threshold 
region. At around the mass of the $\rho$ meson, where the bulk of the 
contribution to $a_\mu^{\rm hard}$ 
comes from,  this correction becomes much smaller 
(see \cite{ecker}). 

The third effect is the difference in the decay widths of 
charged and neutral $\rho$ mesons. Close to the $\rho$ resonance 
this effect can be estimated as:
\be
 \Gamma(\tau^- \to \nu_\tau \pi^0 \pi^-)
\approx \frac{\Gamma_{\rho_0}^2}{\Gamma_{\rho_{-}}^2}
\sigma(e^+e^- \to {\rm hadrons}).
\ee
One can check that in the vicinity of the resonance the 
effect of the  difference in the widths almost cancels the phase space 
corrections due to the difference in the pion masses discussed above.
Let me spell out more precisely how the widths difference is taken 
into account. Starting from the $\tau$ data and taking into account 
all the relevant 
corrections (e.g. the short distance QED correction $S_{\rm ew}$), 
one ends up with the distribution in invariant mass of  $\pi_0 $ and $\pi^-$. 
This distribution is fitted using some parameterization of 
the pion form factor to determine the mass and the width of 
the charged $\rho$. To compute $e^+e^- \to \pi^+ \pi^-$, 
one uses the same parameterization of the pion form factor
but with the mass and the width of 
the neutral $\rho$ instead of similar parameters for the charged one.  

The three effects discussed above are usually taken into 
account in the existing analyses, however, the fourth 
effect, the long distance QED corrections, seems to be more 
problematic. In computing long distance QED corrections one 
usually assumes  that the pions can be treated as  point-like particles. 
This assumption is not quite correct, since in the hard 
renormalization factors $S_{\rm ew}$ 
only  photon virtualities down to the mass of 
the $\tau$ are included. It is quite clear that  photons
with virtualities from the mass of the $\tau$ down to, say,
$1~{\rm GeV}$
certainly resolve the pion and see its quark structure.
The contribution of this momentum region is therefore treated not quite 
correctly in the 
existing estimates. At any rate, the most recent 
calculation \cite{ecker} 
of the long distance QED corrections, performed  using 
scalar QED for {\it point-like pions}, claimed that the long-distance 
QED corrections in $\tau \to \nu_\tau \pi_0 \pi$ add 
$+0.4$ per cent\footnote{Another correction not 
considered in \cite{ecker} but relevant for 
their analysis is the QED correction to leptonic decay mode $\tau
\to \nu_\tau  e \bar \nu_e$ which is used for the normalization 
of the data. Effectively, this correction  \cite{ms} 
adds another $+0.4$ per cent to 
$S^2_{\rm ew}$.}
to the short distance renormalization factor $S^2_{\rm ew}$.

The attitude to the techniques used to obtain this number (chiral power 
counting, point-like pions)
can certainly vary from person to person, 
however, it is important to mention that 
a more complete study of the QED effects, including 
attempts to introduce hadronic  structure, has been performed for 
the decay rate $\tau \to \nu_\tau \pi$ \cite{decker}. 
Since the two processes are similar, it is instructive  
to look at the results in \cite{decker}. Specifically, consider the 
QED corrections to the ratio:
\be
R_{\tau /\pi} = 
\frac{\Gamma(\tau \to \pi \nu_\tau)}{\Gamma(\pi \to \mu \nu_\mu)}.
\ee
The corrections, computed in various approximations,  are \cite{decker}: 
a) short-distance QED\footnote{The short-distance QED corrections both to 
$\tau \to \pi \nu_\tau$ and $\pi \to \mu \nu_\mu$ 
are given by $S_{\rm ew}$. These corrections do not 
cancel out exactly because the appropriate $\mu$ (cf. Eq.(\ref{wils}))
is thought to be different 
in the two processes.}: $-1$ per cent; b) point-like pion (the QED corrections 
to the ratio are finite for  point-like pions): 
$+1$ per cent; c) ``best estimate'' of  \cite{decker} that 
includes hadronic structure and short distance corrections: 
$0.0 \div 0.25$ per cent.
The short-distance QED and the point-like pions are the two extreme 
limits of the problem that 
can be easily understood and I consider their difference as 
the indication that the uncertainty in long distance 
QED corrections to $\tau$ decays can be of order  $1$ per cent.
Other authors (see e.g. \cite{CM}) consider $\pm 0.5$ per cent 
as a more reasonable estimate of this uncertainty.
It is clear that only convincing complete calculation of the QED radiative 
corrections to $\tau \to \nu_\tau \pi^- \pi_0$ can tell us which of the 
two numbers is closer to the truth but 
in the absence of such a calculation I think it makes sense 
to have a conservative attitude.

Finally, we come to an important point that is often not  well 
understood.  Imagine that we have actually succeeded in computing the QED 
corrections to $\tau$ decays and have carefully taken  
into account all the isospin violating effects in transforming 
the $\tau$ data to $e^+e^- \to {\rm hadrons}$. 
Is this the end of the story?  The usual answer here is yes, 
but the correct answer is {\it no}.  This issue is related to 
the discussion above on how corrections due to the widths differences 
of charged and neutral $\rho$ are implemented. Imagine 
that the masses and widths of neutral and charged $\rho$ are obtained 
independently from the fits to $e^+e^-$ and $\tau$ data.
The pion form factor is defined as the $\gamma^* \pi^+ \pi^-$ interaction 
vertex with all the QED interactions between the pions being 
switched off.
Therefore, if one starts from the $\tau$ data, determines the 
pion form factor and uses this form factor with the masses and 
widths for the neutral $\rho$ as obtained from $e^+e^- \to \pi^+\pi^-$,  
one obtains the {\it bare} pion form factor $\gamma^* \to \pi^+ \pi^-$.
One should then compute the final state QED interaction 
corrections and include them into the dispersion integral.
It is also important to stress 
that these corrections are {\it not} included in what is usually
called the next-to-leading order hadronic vacuum polarization corrections
(see Fig.1c). 

How large can these corrections be? To give a simple estimate  
I consider  the $\pi^+ \pi^-$ final state and assume that the  pions are 
point-like particles. In this case the QED corrections are 
easy to compute. The corresponding calculation 
can be found in \cite{schwinger}:
\be
\sigma_{\pi^+ \pi^-} = \frac{\alpha^2 \pi}{s} \beta^3 |F_\pi(s)|^2 
\left [ 1 + \frac{\alpha}{\pi} {\cal F}_{fs}(\beta) \right],
\label{eq16}
\ee
where $\beta = \sqrt{1 - 4m_\pi^2/s}$ and
\ba
&& {\cal F}_{fs}(\beta) = 
\frac {(1+\beta^2)}{\beta} \left \{ 
4 {\rm Li}_2\left ( \frac {1-\beta}{1+\beta} \right )
 +2 {\rm Li}_2 \left (- \frac {1-\beta}{1+\beta} \right )
\right. \nonumber \\
&& \left.
 -3 \log \left ( \frac{2}{1+\beta } \right )\log 
 \left ( \frac {1+\beta}{1-\beta} \right )
-2\log(\beta) \log \left ( \frac{1+\beta}{1-\beta} \right ) 
\right \}
 \nonumber \\
&&  
 -3 \log \left ( \frac {4}{1-\beta^2} \right )  
 - 4 \log(\beta)
 \nonumber \\
&& + \frac {1}{\beta^3} \left ( 
 5 \frac{(1+\beta^2)^2}{4}- 2 \right ) 
\log \left ( \frac { 1+\beta }{1-\beta} \right ) 
+\frac{3}{2} \frac{ ( 1+\beta^2 )}{\beta^2}. 
\label{sqed}
\ea
The radiative correction, as described by this function, is
plotted in Fig.2. First note that the radiative correction 
is rather large; in particular, it is significantly larger 
than the corresponding correction for the production of two 
fermions. Asymptotically, for $\beta \to 1$,   
the correction is $3 \alpha/\pi$ and so one sees a compensation 
of the famous $1/\pi$ factor that is always present in simple 
estimates of the QED radiative corrections. At  threshold, 
the correction is 
again significant because of Coulomb singularity. The bottom 
line is that the correction is relatively large everywhere.

\begin{figure}[htb]
\hspace*{-2mm}
\begin{minipage}{16.cm}
\psfig{figure=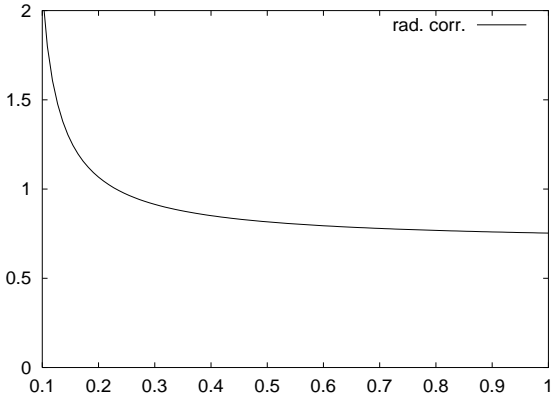,width=80mm}
\end{minipage}
\caption{QED corrections to the production of two pions, 
in per cent, in dependence on $s,~{\rm GeV}^2.$}
\label{fig:light}
\end{figure}

If I use Eq.(\ref{eq16}) in  Eq.(\ref{eq7})
and integrate it  up to $\sqrt{s} = 1$ GeV, I obtain 
the net increase in the contribution of the two pion final state 
to $a_\mu$  by slightly less than  $1$ per cent. This would add
an additional $\sim 50 \times 10^{-11}$ to hadronic vacuum polarization 
contribution to $a_\mu$, if one evaluates it using the $\tau$ data. 

Let me stress that I do not consider the above estimate to be 
an absolute prediction for the missing effect since certainly 
there are  questions here of how well  scalar QED for  
point-like pions actually describes the real world where the pions 
are not point-like and which part of the radiative correction 
is correctly accounted for by modeled (rather than  measured) 
widths difference of charged  and neutral $\rho$ mesons. 
Rather, I think that this estimate should be 
considered as a counter-example to a popular statement that the 
QED corrections are  always ${\cal O}(\alpha/\pi)$ and for this 
reason are  insignificant. 

At the very least, these considerations do imply that the error 
on $a_\mu^{\rm hadr}$ in \cite{davier} is too optimistic. Consider 
the following. The result of \cite{davier} is 
$a_\mu^{\rm hadr}({\rm lo~v.p}) = 
(6924 \pm 56|_{\rm exp} \pm 26|_{\rm th}) \times 10^{-11}$. 
I guess everyone would agree that the long distance QED effects in 
$\tau \to \nu_\tau \pi^- \pi^0$ {\it and} in $e^+e^- \to \pi^+ \pi^-$ 
can be $0.5$ per cent {\it each}. We then get a one percent theory uncertainty 
from the two pion channel and this is $\pm 50 \times 10^{-11}$. Certainly, 
this uncertainty is not completely  taken into account in the theory 
uncertainty  $26 \times 10^{-11}$ from \cite{davier}.
So, at the very least, the systematic uncertainty 
in the result of \cite{davier}, used by the g-2 collaboration in their 
evaluation of the SM result,  is  smaller than it 
should be\footnote{To be fair, I should perhaps say that in many other 
evaluations of hadronic vacuum polarization 
contribution to $a_\mu$ the QED corrections 
are also somewhat forgotten. It is only because of 
the exceptional precision of  the result in \cite{davier} that 
I focus on that reference.}.

The possibility to use the $\tau$ data gave us a useful cross check 
on the accuracy of the $e^+e^-$ data. 
However, it should be clear from the above discussion that the use of the  
$\tau$ data in the analysis of $a_\mu$ requires  essential 
{\it theoretical} input which,   at the required level of precision, 
is hard to justify or check.  It may happen that a better theory 
will convincingly demonstrate that it is possible to control the 
transition from $\tau$ to $e^+e^-$ data with the accuracy well 
below $1$ per cent. In the absence of that, I believe that  the use 
of the $\tau$ data  for computing $a_\mu$
may turn out to be counter-productive. 
As will be seen from the discussion to follow, if one uses 
the $e^+e^-$ data, one at least may try to minimize and   
{\it experimentally}  control certain theoretical assumptions 
necessary to transform the raw data into hadronic vacuum polarization 
contribution to $a_\mu$. I, personally, do not  see how this can be done 
if one starts from the $\tau$ data. 

\section{Hadronic vacuum polarization and the $e^+e^-$ data}

Let me now elaborate on  
the use of the  $e^+e^-$ data to compute the hadronic 
vacuum polarization contribution to $a_\mu$. The point I would like 
to make here is that the $e^+e^-$ data offers a relatively clean and, 
what is perhaps more important, verifiable approach to evaluating 
hadronic vacuum polarization with the required precision.

There are many papers where the $a_\mu^{\rm hadr}({\rm lo~v.p})$ 
is evaluated from $e^+e^-$ data and I will not 
discuss all of them (for the recent reviews and calculations 
see \cite{j2000,mlr,narrison}).

For the purpose of the illustration I 
will use the result of \cite{ej1995}\footnote{The renormalization group 
improved result quoted in 
\cite{ej1995} as the principal result can not be used together with 
higher order QED corrections to vacuum polarization contribution
computed in \cite{krause}. For this reason, I quote below 
the result in \cite{ej1995} that is obtained {\it without} renormalization 
group improvement.} as well as more recent 
updates by one of these authors\cite{j2000}. 
Their results are:
\ba
a_\mu^{\rm hadr} ({\rm lo~v.p})~~~  && =
7025(150) \times 10^{-11}~~~[1995],
\nonumber \\
&& = 
6974(105) \times 10^{-11}~~~[2000].
\label{ej}
\ea

By comparing these numbers to the $\tau$-based 
result $6924(62) \times 10^{-11}$  \cite{davier} that has been used 
by the g-2 collaboration for the calculation of the SM prediction 
for the muon anomalous magnetic moment, one sees that the $e^+e^-$ 
numbers are somewhat larger. However, 
it is not simply $\tau$ vs. $e^+e^-$ that determines 
the difference of the two results. 
   
To show this, let me note that there exists a dedicated analysis of the 
CVC hypothesis for various exclusive channels \cite{ivanchenko}
based on comparison of the $e^+e^-$ and $\tau$ data. 
For example,  if one uses the $e^+e^-$ data and CVC to predict 
the corresponding  branching ratio $B(\tau^- \to \pi^0 \pi^- \nu_\tau)$, one 
obtains \cite{ivanchenko}:
\ba
&&B(\tau^- \to \pi^0 \pi^- \nu_\tau)|_{\rm CVC} = 24.52 \pm 0.33,
\nonumber \\
&&B(\tau^- \to \pi^0 \pi^- \nu_\tau)|_{\tau~{\rm data}} = 25.32 \pm 0.15,
\ea
which implies that the $\tau$ data actually predicts a {\it larger} 
contribution of the $\rho$ resonance, an opposite situation to 
what one sees in the final numbers for $a_\mu^{\rm hadr}$. The explanation 
for that is that the calculation  \cite{davier}  is 
more than just the $\tau$ data; it also involves e.g. the use 
of perturbative QCD down to $1.8~{\rm GeV}$; the approach 
the authors of \cite{ej1995}  try to avoid. Since the old 
$e^+e^-$ data at around 
$2~{\rm GeV}$ is {\it significantly higher} than the pQCD results at 
that energies, this turns out to be an important  part of the 
difference {\it in the central values} of \cite{davier} and 
\cite{ej1995}.

Let me now say a few words about the composition of the errors
in $e^+e^-$ data.  The errors are distributed as \cite{ej1995}:
\ba
2 m_\pi < E < 0.8~{\rm GeV}:~~~~~~~ && \sim 100 \times 10^{-11}, 
\nonumber \\
 0.8~{\rm GeV} < E < 1.41~{\rm GeV}:~~~~~
&&  \sim 50 \times 10^{-11},
\nonumber \\
 1.41~~{\rm GeV} < E < 3.10~{\rm GeV}:~~~~~
&&  \sim 50 \times 10^{-11}.
\nonumber
\ea
Other errors seem to be negligible. 

Another important benchmark \cite{jrc98}
is how accurate different exclusive channels
should be known if the final result for $a_\mu$ is to be known with the 
precision $100 \times 10^{-11}$ (I assume that the errors from 
individual channels are combined in quadrature).
The $e^+e^- \to \pi^+ \pi^-$ final 
state should be known at the level of one per cent; $\omega \to 3 \pi$, 
$\phi$, $4\pi$ and the contribution from above $2~{\rm GeV}$ 
should be known roughly at the  $10$ per cent level. 
 To decrease the  error to something like 
$30 \times 10^{-11}$ all these uncertainties should 
be scaled down by a factor of three, approximately.

Some of these errors, most notably  the error from the region 
below $\sqrt{s} \sim 1~{\rm GeV}$ will go down 
significantly due to new data from the  VEPP-2M collider at Novosibirsk. 
Their final results  are not made public yet, however the anticipated 
accuracy of $0.6$ per cent is already known.  The other 
improvement will, potentially, come from BEPC \cite{zhao}. They are measuring 
the value of $R(s)$ at $\sqrt{s} > 2~{\rm GeV}$. At energy 
regions below $J/\psi$, their preliminary 
results are accurate to within $7$ per cent and  they are about $15$ 
per cent lower than the earlier results of Mark~I and Gamma2 experiments 
and a bit higher than the pQCD results. 

Let us now return to the two pion channel and the forthcoming Novosibirsk
results. Since the $0.6$ per cent
accuracy is outstanding, it is important 
perhaps to spell out some delicate issues that might help to make it 
more believable.

The major point to realize about the  difference 
in the use of $e^+e^-$ data for the evaluation of $a_\mu$ as compared 
to $\tau$ data is that in principle when one uses the 
$e^+e^-$ data one needs much less  theoretical input.
Clearly, there are certain things to be worked out, like 
QED corrections related to initial state radiation and 
vacuum polarization, but the part of the QED corrections 
that describes the interaction of $\pi^+$ and $\pi^-$ in the 
final state is already in the data. Still there remains a 
potential problem  that I discuss below.
It is important to distinguish at this point the pion form factor 
as used for comparison with different models and for the 
determination of the mass and the width of the $\rho$ meson 
where, by definition, 
the final state QED radiative corrections are not included and 
the cross section $\gamma^* \to \pi^+ \pi^-$ for the purpose 
of $a_\mu$ calculation, where the final state QED radiative corrections 
should be kept intact. I believe  that the two pion channel 
data from CMD2 will be analyzed this way \cite{private}. 

Although appreciating the difference 
between $F_\pi$ and $\gamma^* \to \pi^+\pi^-$ when the QED
effects are considered is important, it is equally important 
to realize potential problems with the Novosibirsk analysis. 
First of all, the Monte Carlo event generator for the two pion 
channel that is used 
to analyze the data is  based on  point-like pions \cite{kuraev}. 
This might be  a potential limitation.

Another problem is that the experimental analysis starts with imposing certain 
cuts to isolate {\it two pion} final state. The major requirement 
here is that the two pions are essentially  back-to-back and 
therefore this cut excludes the $\gamma \pi^+ \pi^-$ final 
state where the photon is radiated off one of the pions or an electron or 
positron at a relatively large angle. 
My estimates\footnote{I used the Monte Carlo event 
generator  for $e^+e^- \to \pi^+ \pi^- \gamma$ \cite{binner}; I am grateful to 
G. Venanzoni for help with that.} show that 
the cuts applied in the recent CMD2 measurement of $e^+e^- \to \pi^+ \pi^-$
\cite{binp10} remove from $50$ to $80$ per cent of  
$e^+e^- \to \pi^+ \pi^- \gamma$. The degree of rejection is 
energy dependent (for smaller $\sqrt{s}$ fewer events are 
rejected, since the transverse momentum of photons is smaller)
and it is also different for final and initial state radiation.
At any rate, there are events that are rejected right up front by 
experimental cuts 
and that, potentially, should be put back since no independent 
measurement of $\pi^+ \pi^- \gamma$ for large angle photons 
at the energy region around  the $\rho$ meson 
has been reported so far. The only way 
this can be done without doing the measurement of $\pi^+ \pi^- \gamma$ 
is to use the Monte Carlo. One should realize, however, that in this 
way one puts back the large angle photon emission by using point-like 
pions and this is the most problematic region for the point-like pion 
approximation to begin with.

To see the importance of higher order QED corrections to the  hadronic 
vacuum polarization contribution, one can can either recall the 
discussion of Schwinger's correction to the two pion channel 
in the previous Section or look at  a much cleaner next-to-leading 
order hadronic vacuum polarization calculation \cite{krause} (see Fig.1b). 
Take any diagram 
that contributes to the two-loop QED correction to g-2 
and insert the hadronic vacuum polarization in to either of the two virtual 
photon lines. This gives a correction 
$\Delta a_\mu^{\rm had}({\rm nlo~v.p.}) = 
-100(6) \times 10^{-11}$ \cite{krause}. One particular contribution 
to this number comes from combined  leptonic and hadronic vacuum
polarization in the one-loop diagram.  This one is interesting  since,
in some sense,   it is related to the vacuum polarization insertions
in $e^+e^- \to {\rm hadrons}$. This  contribution is about 
$100 \times 10^{-11}$ and one clearly sees how large the corresponding 
corrections can be.

Let me now discuss the question of what should be done 
in order to ensure a careful job on 
$e^+e^- \to \pi^+ \pi^-$.  The best thing, of course, 
is if one actually measures the $\pi^+ \pi^- \gamma$ channel 
separately and checks that it actually matches the $\pi^+ \pi^-$ 
channel as far as the acolliniarity angle of the two pions is 
concerned. This might be quite a tough measurement since at the 
very end one will have to disentangle the large angle final state radiation 
from the initial one.

However, if  relatively energetic photons and pions are detected,
one can make a study of the charge asymmetry of the produced pions 
\cite{binner}. In case when the hard photon is tagged, this effect comes 
from the interference between initial and final state radiation and 
is therefore linear in 
the final state radiation amplitude. Thus, the charge asymmetry of the 
produced pions in $\pi^+ \pi^- \gamma$ events with all the particles
emitted at relatively large angles, gives a direct handle on the 
amplitude of the final state radiation. I believe these kind of 
studies should be done to cross check the {\it model} (and, certainly, 
the scalar QED for the interaction of pions with photons {\it is} 
a model) used in the Monte Carlo event generators.  There are some 
preliminary results from DAPHNE \cite{achim} and also old results 
from Novosibirsk on $\pi^+ \pi^- \gamma$ channel \cite{dol}
that seem to indicate that the 
point-like pion approximation 
works amazingly well in the energy range  around
$1~{\rm GeV}$; however it is still not completely conclusive.

At any rate, summarizing the use of $e^+e^-$ data for $a_\mu$ predictions, 
I can say that, currently, the $e^+e^-$-based evaluations 
of  $a_\mu^{\rm hadr}$ have a somewhat higher central value and 
larger error bars than the value of $a_\mu^{\rm hadr}$ \cite{davier} 
used in \cite{g-2recent}
which implies that if one evaluates the SM prediction for $a_\mu$ 
using the $e^+e^-$ data, the g-2 ``crisis'' becomes less acute.
The precision of $e^+e^-$-based evaluations will improve once 
the new data from the low energy $e^+e^-$ machines is incorporated.
It is important to realize that at this new level of precision 
new questions, primarily related to the correct treatment of QED radiative 
corrections, will start to appear. However, it seems to me
that a program of measurements and analysis 
can be set up that makes it possible to control every step on the way 
from the $e^+e^-$ data to the muon anomalous  magnetic moment. 
This is the principal  difference with the $\tau$ data. 

\section{ The light-by-light scattering contribution}

The light-by-light is probably the most tricky thing in the muon 
g-2 calculation. The trouble is that, in contrast to hadronic vacuum
polarization, there is no simple way to relate this contribution 
to anything  observable. In this situation, one has to resort 
to  models to describe low-energy hadron dynamics and 
then the question of the reliability of a certain model 
becomes central.

Before going into the discussion of the {\it delicate} issues related 
to hadronic light-by-light scattering, let me first clarify a misconception 
that, as it seems to me, is quite common in the current literature.
The issue I want to address is what is the relevant scale for the 
loop momenta that determines the contribution of the light-by-light
scattering diagrams to $a_\mu$. It is usually said in the literature that 
this is the mass of the muon; for hadronic light-by-light, this statement 
is not correct. 

To see this, it is useful to analyze a simple QED example by computing the 
light-by-light scattering contribution of the fermion of the mass 
$M$ to the anomalous magnetic  
moment of the muon. Lets introduce the variable $x = M/m_\mu$ and consider
the limit $M \gg m_\mu$ which is relevant for hadronic light-by-light
(both the mass of the pion and the constituent quark masses are larger 
than the mass of the muon). The result of the QED calculation is then
\cite{rl}:
\ba
a_\mu |_{x \gg 1} = && \left( \frac{\alpha}{\pi} \right )^3 
\Big \{ \frac {0.615}{x^2}  +
 \frac {1}{x^4} \left (
- 0.2~\log^2 x 
\right. \nonumber \\
&&  \left. - 0.33 \log x  - 0.15 \right ) + {\cal O}(x^{-6}) \Big \}.
\label{xg1}
\ea
The question I would like to discuss is how the above result 
can be obtained using either the effective field theory technique
or, equivalently,  asymptotic expansion 
of the relevant Feynman diagrams in $m/M$.
  
Upon examination, one can easily identify three expansion 
regimes for a generic light-by-light scattering 
diagram. The first regime is the Taylor expansion of the diagram 
in the ratio of $m_\mu/M$. In this regime, the momenta of all three virtual 
photons are of the order of the {\it heavy} fermion mass. In the language 
of effective  field theories, this is the contribution that directly 
induces the anomalous magnetic moment operator in the effective Lagrangian:
\be
{\cal L}_1 = c_1 
\frac{m_\mu}{M^2}~\bar \psi \sigma_{\mu \nu} \psi F^{\mu \nu},
\ee
where $c_1$ is some constant.

The second expansion regime is related to Euler-Heisenberg Lagrangian
for photons. In this case, the momenta of all three virtual 
photons are of the order of the  muon mass $m_\mu$ and the corresponding 
part of the effective Lagrangian is:
\be
{\cal L}_2 =  \frac{\alpha^2}{360~M^4}  \left [
4 \left ( F^{\mu \nu} F_{\mu \nu} \right )^2 
+ 7 \left (\frac{1}{2} 
\epsilon^{\mu \nu \alpha \beta} F_{\mu \nu} F_{\alpha \beta} 
\right )^2
\right ].
\ee
This piece in the effective Lagrangian determines the strength 
of the low-energy photon-photon scattering.

The third expansion regime is the following: one of three virtual 
photons has the momentum of order $m_\mu$, while  two others have large 
$\sim M$ virtualities. Integrating out 
heavy degrees of freedom in this configuration 
induces the following term in the effective Lagrangian
\be
{\cal L}_3 =  \frac{1}{M^4} 
F_{\mu \nu}  F_{\alpha \beta}  \bar \psi \left [ 
m_\mu \Gamma_1^{\mu \nu \alpha \beta}
+ \Gamma_2^{\mu \nu \alpha \beta \rho} D_\rho
 \right] 
\psi.
\ee
Here $D_\rho$ is the (QED) covariant derivative and $\Gamma_{1,2}$ 
are appropriate Lorentz tensors; their exact form is of no concern 
to us here. The Lagrangian ${\cal L}_3$, in principle, contributes to 
low energy muon-photon scattering. 

An important point to notice is that only the {\it first} 
expansion regime (and therefore the effective operator ${\cal L}_1$ )
gives an ${\cal O}(M^{-2})$ contribution to the anomalous magnetic moment,
whereas the other two  {\it only} start to contribute at ${\cal O}(M^{-4})$. 
 This trivial observation immediately implies that,
no matter how small the muon mass is, {\it there is no low energy 
information one can use to determine the leading contribution 
to the muon anomalous magnetic moment coming from  large momentum 
scales}. The only thing one can do (and, very roughly,  
this is what one usually does) 
is to still use the low energy effective Lagrangians ${\cal L}_2$ 
and ${\cal L}_3$ and compute the Feynman diagrams using hard 
momentum cut-off $\lambda$. If one takes this cut-off to be 
$\lambda \sim M$, one generates the contributions 
$m_\mu^2 \lambda^2/M^4 
\sim m_\mu^2/M^2$ even from higher dimensional operators;
these contributions, however, have nothing to do with the correct result.

We therefore see that the typical scale for the loop momenta in 
hadronic light-by-light is set by the hadronic scale and not 
by the muon mass. Since the lightest hadron is the pion and 
since it contributes to the light-by-light diagrams, one can hope that 
the leading contribution  is determined by momentum transfers of  
order $m_\pi$ and  this is  small as compared to 
the scale of chiral symmetry breaking $\sim 1~{\rm GeV}$.
If this is true, this seems to exclude the possibility
to use quarks and gluons as the relevant degrees of freedom for this 
calculation and, instead, one can argue that 
for the momentum transfer being that small,  chiral perturbation 
theory should be quite reliable.  For this reason, all the calculations 
of hadronic light-by-light  scattering  are based on chiral perturbation 
theory  gauged with $U(1)$ electromagnetism. However, it turns out 
that the integrals {\it do not} converge at momentum scales around $m_\pi$ 
and for this reason  chiral perturbation theory  {\rm alone} 
can not   produce a definite prediction. 
One then resorts to phenomenological  models such as large $N_c$, 
extended  Nambu-Jona-Lasinio, hidden local symmetry etc.

Within this framework, three major contributions to light-by-light 
scattering are distinguished by using large $N_c$ and chiral power counting.
The first is the box of charged pions, 
the  second is the contribution of the neutral pseudoscalar boson through 
a transition $\gamma^* \gamma^* \to {\rm P} \to \gamma^* \gamma$ where 
$P = \pi^0, \eta,\eta'$.
The third contribution is due to  constituent quark loops.

The simplest calculation to be done is to compute the contribution 
of the quark loop in QED and a similar contribution with elementary 
pions in scalar QED. In this way one gets \cite{klbl}:
\ba
&& a^{\rm hadr}_\mu({\rm lbl,~pions}) = -44.58(23) \times 10^{-11},
\label{pionloop}\\
&& a^{\rm hadr}_\mu({\rm lbl,~quarks}) = 62(3) \times 10^{-11},
\label{quarkloop}
\ea
where the constituent masses for 
light quarks ($m_u = m_d = 0.3~{\rm GeV}$ and $m_s = 0.5~{\rm GeV}$)
have been used. One sees that the two contributions 
in no way match {\it since they differ by a sign}.

To improve on that result  one would like to take into account 
a) the interactions between pions or quarks and b) the modifications of the  
pion-photon coupling for the off-shell photons.
The self-interaction of pions is thought to be described by higher-derivative 
terms in the chiral Lagrangian. Generically, 
these terms generate corrections 
of the form $ p_{\rm typ}^2/\Lambda_{\chi}^2$, where $p_{\rm typ}$ is the 
typical pion momenta and $\Lambda_{\chi} \sim 1~{\rm GeV}$ is the 
scale of the chiral symmetry breaking in three flavor QCD. Taking 
$p_{\rm typ} \sim m_\pi$ for the estimate, one expects $\sim 1$ per cent
corrections but I doubt that this estimate is actually correct (see below).
The trouble is that if one wants to go from the estimate to 
the calculation one would not be able to produce an unambiguous 
answer both within chiral perturbation theory {\it and} the models
used currently for the calculation of the hadronic light-by-light.
For this reason the problem of self interactions between pions and quarks 
is  ignored in the current literature.

Let us now consider the modification of the pion to photon coupling for 
the off shell photons. The simplest model is 
the vector-meson-dominance (VMD) model, 
that essentially postulates that photons interact with pions through 
a transition to the $\rho$ (or any vector) 
meson. The propagator of the photon is  then
modified to be:
\be
\frac {i}{q^2} \to \frac{i m_\rho^2}{q^2 \left (m_\rho^2 - q^2 \right ) }.
\ee
How much do we expect the free pion result would change 
once the VMD modification of the  photon propagator is introduced? 
If all the integrals converge at around
$m_\pi$, as is usually assumed, 
one would expect the modification to be 
of the order $m_{\pi}^2/m_\rho^2 \sim 0.04$. In reality, the 
situation is quite different. First of all, there {\it is} a significant 
modification --   the result with VMD is smaller by a factor $6 -10$. 
This implies that original integrals actually converge at momentum 
scales {\it much higher} than the mass of the 
pion (what about chiral perturbation 
theory in such a  situation?).

There is a good illustration of this fact in \cite{klbl}.  
In the complete calculation one may try to consider the masses 
of the pion and the $\rho$ meson as variable quantities and 
ask about asymptotic behavior of the results once these masses
are changed. For example, by keeping the mass of the $\rho$ meson fixed 
but changing the mass of the pion, one obtains the $m_\pi^{-2}$ 
scaling law; this indicates that the integral over the pion box 
subdiagram is saturated at small momenta. On the contrary, 
changing the mass of the $\rho$ meson gives the asymptotic behavior:
$$
a^{\rm hadr}_\mu({\rm lbl}~m_\pi,m_\rho) 
- a_\mu({\rm lbl}~m_\pi,\infty) = 0.23 
\left (\frac {m_\mu}{m_\rho} \right ) \left ( \frac {\alpha}{\pi} \right )^3.
$$
This  shows that this contribution is quite large\footnote{
This  linear dependence might be an artifact of the fitting 
procedure. Using the data in \cite{klbl} I can fit the pion 
box contribution to  $a_\mu(m_\pi,m_\rho) - a_\mu(m_\pi,\infty) = 
(\alpha/\pi)^3~0.8~m_\mu^2/m_\rho^2 \log^2 m_\rho/m_\mu$ which is
more consistent with what one should expect from the Feynman 
integrals with two largely different scales involved.}.  
The final result for this contribution quoted in \cite{klbl} is:
\ba
&& a^{\rm hadr}_\mu({\rm lbl,~ pions}) 
=  \left ( -0.03557 + 0.23 \frac{m_\mu}{m_\rho} \right ) 
 \left ( \frac {\alpha}{\pi}\right )^3 = 
\nonumber \\
&&~~~~~~~~~~~~~~-0.00355  \left ( \frac {\alpha}{\pi}\right )^3 =
-4.5(8.1) \times 10^{-11}.
\label{pions}
\ea
a reduction by one order in magnitude as compared to the point-like pion 
result Eq.(\ref{pionloop}). 

A very similar pattern is observed if the  VMD modification is 
applied to  photon couplings to  constituent quarks.  
The final result quoted for this contribution 
in \cite{klbl} is 
\be
a^{\rm hadr}_\mu\left ( {\rm lbl,~quarks} \right ) = 9.7(11.1) \times 10^{-11}.
\label{quarks}
\ee

The last  contribution to be considered is the one from the 
pseudoscalar meson pole (see Fig.3) 
and, in view of the smallness of the pion and 
quark  contributions (after VMD ), it turns 
out to be the dominant one. The result is 
\cite{klbl2}: 
\be
a^{\rm hadr}_\mu({\rm lbl,~pole}) = -82.7(6.4) \times 10^{-11}.
\label{pole}
\ee
Here, $\pi^0$, $\eta$ and $\eta'$ are contributions are
taken into account. The $\pi^0$ contribution is about $70$ per cent,
with the rest being distributed equally between $\eta$ and $\eta'$.
Large contributions from $\eta$ and $\eta'$ look surprising and I comment 
on it at the end of this Section.
The result in Eq.(\ref{pole}) is obtained 
by using some constraints on the $\gamma^* \gamma^* \pi^0$ interaction
vertex from the measurement of the pion transition 
form factor by the CLEO collaboration
\cite{cleo}.

The final result quoted in \cite{klbl2} is obtained by summing 
up Eqs.(\ref{pions},\ref{quarks},\ref{pole}) and adding to it 
small ($1.7 \times 10^{-11}$) axial-vector meson contribution.
The result reads:
\be
a_\mu({\rm lbl,total}) = ( -79 \pm 15.4 ) \times 10^{-11},
\label{kinf}
\ee
if the errors of the individual contributions are added in quadratures.
If they are added linearly, the error on this contribution is 
$\pm 25 \times 10^{-11}$.

Another calculation of the hadronic light-by-light contribution,
based on   Extented Nambu-Jona-Lasinio model,
has been presented in \cite{bijnes}. 
The result quoted in \cite{bijnes} is\footnote{The calculation 
of Ref.\cite{bijnes} is performed with the cut off 
on the loop momentum and the sensitivity of the final result 
to the cut off is studied. The authors observe that
in order to have a stable result 
for some  of the contributions, the integration should be extended up 
to several GeV. The error of the final result in \cite{bijnes} 
is obtained if the errors on individual contributions 
are added linearly.}:
\be
a_\mu({\rm lbl,total}) = ( -92 \pm 32  ) \times 10^{-11}
\ee
and is therefore close to the result in Eq.(\ref{kinf})
(see Ref.\cite{bijnens1} for the comparison of 
\cite{bijnes} and \cite{klbl2}).

The final result for the light-by-light scattering contribution 
used by the g-2 collaboration is the arithmetic average of the two 
results; a similar procedure is applied to calculate the uncertainty. 
Even if the two results are to be trusted, a sensible thing to do is, perhaps,
to take the uncertainty to be so large as to cover the whole range 
as allowed  by individual results. We then  have:
\be
a_\mu({\rm lbl,total}) = -85(38) \times 10^{-11},
\ee
and it is important to stress that the only thing this uncertainty 
is supposed to represent is the {\it model dependence}. 

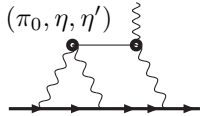
\begin{figure}
\begin{center}
\hfill
\begin{picture}(120,40)(0,0)
 \SetScale{.8}
 \SetWidth{1.5}
 \Line(20,20)(110,20)
 \SetWidth{0.8}
  \ArrowLine(25,20)(42,20)
 \ArrowLine(42,20)(62,20)
 \ArrowLine(67,20)(87,20)
 \ArrowLine(80,20)(96,20)
 \ArrowLine(100,20)(110,20)
 \SetWidth{0.5}
  \Photon(35,20)(50,50){1.8}{4}
  \Photon(65,20)(50,50){1.8}{4}
  \Line(50,50)(80,50)
  \SetWidth{2} 
  \Oval(50,50)(2,2)(0)
  \Oval(80,50)(2,2)(0)
 \SetWidth{0.5}
  \Photon(80,50)(95,20){1.8}{4}
  \Photon(80,50)(80,70){1.8}{4}
 \Text(35,55)[t]{($\pi_0,\eta,\eta'$)}
\end{picture}
\hfill\null\\
\SetScale{1}
\end{center}
\caption{The pseudoscalar contribution to g-2.}
\label{fig:examplediagram1}
\end{figure}

In view of the large value of the pseudoscalar pole contribution 
it is appropriate to discuss it in some detail. Again, it is instructive 
to ask the question about  momentum flows in the corresponding diagrams
to see potential troubles. The $\pi^0 \gamma \gamma$ vertex 
for the on-shell pion and photons is given by:
\be
{\cal L} = -\frac{\alpha}{8\pi f_\pi} \pi^0 \ep^{\mu \nu \alpha \beta} 
F_{\mu \nu} F_{\alpha \beta}. 
\ee
Considering this as a new vertex in the low-energy effective Lagrangian
and inserting it into the diagram that describes its contribution 
to the anomalous magnetic moment of the muon, we observe that the result 
is divergent and this divergence is cut off by the pion 
transition form factor. It is easy to work out the corresponding contribution 
in the leading logarithmic approximation with the result:
\be
a^{\rm hadr}_\mu ({\rm lbl,}~\pi_0) \sim 
\left( \frac{\alpha}{\pi} \right )^3 
\frac{m_\mu^2}{(4 \pi f_\pi)^2} 
\int \limits_{m_\pi}^{m_\rho} \frac{{\rm d}k}{k} \log \frac{m_\rho}{k},
\label{pi0}
\ee
where $k$ is the momentum that runs along the $\pi_0$ line. In doing 
this estimate I assumed that the integrals are cut off from above by the 
hadronic scale comparable to the mass of the $\rho$ meson. 

Eq.(\ref{pi0}) gives us useful information about the structure 
of the divergences and, hence, about momentum flow. 
First, the divergences are double logarithmic 
and not just single logarithmic, as it is sometimes claimed in the 
literature. Second, there are two divergences. One is associated 
with large virtualities of the photons, when the momentum that goes through 
the pion line is kept fixed. The second one is associated with large 
virtualities of the {\it pion}. 

Currently, both of these divergences are cut off by adopting VMD prescription 
for the photon lines. This approach is supported by the phenomenological 
success of the VMD models in describing 
the $\gamma \gamma^* \to \pi^0$ transition 
form factor.  Also, using this kind of the regularization one can 
compute the decay width $\pi^0 \to e^+e^-$ and similar  and obtain 
a reasonable agreement with the data.  For this reason it probably
makes complete sense to use the VMD motivated regularization for 
the photon loop subdivergence. It is less clear if the same regularization 
makes any sense for the other divergence associated with highly 
virtual $\pi^0$. The point is that this kinematics is not 
related to any observable form factor and it is not clear what 
the highly virtual $\pi_0$ means. Ideally,  this configuration 
should somehow match on to the quark box QCD diagram, but no one 
knows how to implement that in practice.

What also seems rather intriguing is the fact that the contributions 
of $\eta$ and $\eta'$ are quite large. Approximately, they are 
one fourth of the $\pi_0$ contribution
in spite of huge difference in  masses. For the sake of the 
argument, consider $\eta$ meson. Its coupling to two photons is roughly 
the same as the $\pi^0 \gamma \gamma$ coupling. 
Taking into account that the ratio of the masses is 
$m_\pi/m_\eta \sim 1/4$, one concludes that the suppression from the 
loop integral goes like $1/m$ when the mass is increased. A similar 
conclusion is reached when the $\eta'$ contribution is analyzed.

Another point I would like to mention is that it is unclear if 
the quark loop contribution should be damped by introducing 
VMD modification of the photon propagator.  Imagine that 
we want to set up a calculation in the effective field theory 
framework. To do that, we are supposed to introduce a factorization scale. 
Above this scale, we do the calculation with quarks
and gluons and below this scale with hadrons. If one looks
at the calculation from this perspective, then the calculation 
of the quark contribution to light-by-light scattering 
should be cut off in the infra-red; for technical reasons this is 
achieved by introducing the quark masses $\sim 300~{\rm MeV}$. 
On the other hand, the calculation with hadrons is regularized using VMD 
modification for the photon propagators which cuts off the 
integrals {\it from above}. In spite of the fact that the 
cut off is implemented differently in two parts of the calculation, 
in my opinion, the set up described above is internally consistent. 
On the other hand, it shows that the introduction of the VMD modified 
photon propagators into the quark contribution 
to light-by-light scattering  is  not quite logical.  
The quark contribution is supposed to describe physics at energy 
scales $2m_Q \le k$. When, {\it in addition to quark masses}, 
the VMD is introduced into loop integrals, the integration is cut 
off at $k \le m_\rho \sim 2m_Q$ and this is clearly {\it outside}
the momentum range that the quark contributions is supposed to 
cover. For this reason, it seems to me that the quark contribution 
Eq.(\ref{quarkloop}) {\it without} VMD suppression might be a more 
appropriate description for the contribution of the high energy region.
If so, the result for hadronic light-by-light scattering 
might  receive additional positive contribution. 

Let me finally comment on the argument used to justify the 
application of the Extended Nambu-Jona- Lasinio model to the calculation 
of the hadronic light-by-light contribution to g-2.
A possible  check is to apply the same model to the calculation of 
hadronic vacuum polarization contribution to g-2 and 
see how well the result based on  experimental data can be reproduced.
This has been done in Ref.\cite{rafael} where the claim is that 
within the class of models like the ones used in \cite{klbl,bijnes},
the hadronic vacuum polarization contribution can be predicted to 
within $15$ per cent.  This fact, by itself, is nice but I am not 
sure how restrictive it is.  Let me take three constituent quarks 
with masses $200~{\rm MeV}$ and use the lowest order cross section
for $e^+e^- \to q \bar q$ to compute the ``hadronic'' vacuum polarization 
to $a_\mu$. Including only $u,d$ and $s$ quarks and integrating up to 
$2~{\rm GeV}$,  I get the  hadronic vacuum polarization contribution 
to $a_\mu$ to be $\sim 5000 \times 10^{-11}$, a perfectly reasonable number. 
Using the same ``model'' for computing hadronic light-by-light, 
I would have obtained the result close to $100 \times 10^{-11}$. 
This number  is in the  ``correct'' range but the {\it sign} is opposite. 
I think this shows my point -- in light-by-light we are sensitive to 
much more detailed structure of the hadronic interactions than we can 
check using hadronic  vacuum polarization and this fact 
seem to matter after all.

To summarize my discussion of hadronic light-by-light, I would like 
to stress that the major question here is  the model dependence of the result 
and in this respect, in my opinion, the agreement between the two 
independent calculations \cite{klbl,bijnes} does not tell us much
since the models used in these calculations are similar. As I have 
discussed above in detail, the momentum scales that control the 
hadronic light-by-light scattering contribution are neither the mass 
of the muon nor the mass of the pion, as is often assumed. For this reason,
any low energy hadronic model that is used for such a calculation, 
should be accurate up to $\sim 1~{\rm GeV}$ and, as far as I understand, 
there are not too many such models on the market.  It would certainly 
be very helpful if this problem would come under scrutiny of the 
low-energy hadron physics community.

\section{Conclusions and future prospects}

What can be expected for the muon anomalous magnetic moment 
in the future? The g-2 collaboration 
will improve the accuracy of their result to $40 \times 10^{-11}$.
However,   even with this accuracy 
the interpretation of this result  will depend on our ability 
to estimate the hadronic contribution to  g-2. 

The analysis of $e^+e^- \to {\rm hadrons}$ from Novosibirsk is 
in its final phase. This implies that soon the new $e^+e^-$ 
based estimate of the hadronic vacuum polarization will be 
available. Hopefully, it will include a proper treatment of 
QED radiative corrections. With this new result, 
there will, probably, be no need to use the $\tau$ data, since 
the $e^+e^-$ data will become sufficiently accurate. 

The value of $R(s)$ will probably be re-measured by using 
radiative return by KLOE, CLEO and BaBar collaborations
at existing facilities. It is hard to imagine that the radiative 
return based measurements will achieve a one per cent accuracy;
$3-5$ per cent accuracy is, probably, within reach. This might 
be sufficient for the energy region above $1~{\rm GeV}$ but it is 
not sufficient for the $2 \pi$ channel.
Therefore, to a large extent, 
the $e^+e^-$ based interpretation of the muon anomaly,  will 
hang on the new Novosibirsk data.

On the theory side, the four loop QED radiative corrections 
were not checked by an independent calculation. Certainly, 
these kinds of calculations are much less rewarding than 
model building and so it seems that the chances to have 
any progress here are slim.

Finally, the real bottleneck seems to be the hadronic light-by-light 
scattering contribution, because all the existing arguments
that make one believe in the validity of theoretical 
estimates are, from my viewpoint, rather inconclusive.
There is talk about getting some help from lattice field theory
but it is difficult to believe in that. It would be of great help
if the people who do low energy hadron physics phenomenology would 
come up with a {\it radically} 
different model (as compared to what is used now) 
to do the calculation of hadronic  light-by-light. 
If this happens, there will be at least some indication on how large  
the real model dependence is.

From what I have said, it should be quite clear that so far there is
really {\it no} g-2 crisis. For the purpose of illustration, consider 
the following estimate. Let us 
take the recent $e^+e^-$ based re-evaluation of the hadronic vacuum
polarization \cite{j2000},
which central value is about $50 \times 10^{-11}$ higher
than the result in \cite{davier}.
Let me also use $38 \times 10^{-11}$ as an uncertainty 
in the light-by-light. Finally, let me add all the theory errors linearly.
I obtain:
\be
\left [ a_\mu^{\rm exp} - a_\mu^{\rm th} \right ]\times 10^{11} 
= 377 \pm 150 |_{\rm exp} \pm 156|_{\rm th}.
\label{new}
\ee
Clearly, if one presents it in such a way there is no significant 
discrepancy. Again, to avoid misunderstanding, let me stress that I do not 
consider Eq.(\ref{new}) as my ``best'' estimate of the current 
difference between  the theory and experiment; 
rather, Eq.(\ref{new}) shows that the theory uncertainty on 
the SM value of $a_\mu$, if evaluated conservatively, is significant
and  a great deal of work will be required to reduce it 
to the level required by experiment.

{\bf Acknowledgments:} This work was supported in part by DOE under 
grant number DE-AC03-76SF00515.  Originally these notes have been prepared 
for the seminar at SLAC Theory Group on the muon anomalous magnetic moment.
I acknowledge the effort by Sidney Drell and Burton Richter to convince me 
to make these notes public. 
I am very grateful to Stanley Brodsky, 
Andrzej Czarnecki, Lance Dixon, Sidney Drell,  Simon Eidelman,  Johann K\"uhn, 
William Marciano, Michael Peskin, 
Burton Richter and  Marvin Weinstein  for stimulating conversations.

\end{document}